\input amssym.def
\input amssym.tex

\def\Z{\Bbb Z}
\def\hfb{\hfill\break}

\input harvmac
\magnification 1200 
\hsize=16.5truecm 
\vsize=23.5truecm 
\overfullrule=0pt
\parskip=3pt
 at 14truept
\overfullrule 0pt
\voffset -10pt 


\def\hexnumber@#1{\ifcase#1 0\or1\or2\or3\or4\or5\or6\or7\or8\or9\or
        A\or B\or C\or D\or E\or F\fi }

%
\def\boxit#1{\leavevmode\kern5pt\hbox{
	\vrule width.2pt\vtop{\vbox{\hrule height.2pt\kern5pt
        \hbox{\kern5pt{#1}\kern5pt}}
      \kern5pt\hrule height.2pt}\vrule width.2pt}\kern5pt}



\def\\{\hfil\break}
\def\la{\lambda}

\def\al{\alpha}
\def\be{\beta}

\def\b#1{\big(#1\big)}

\def\+{\oplus}

\def\b#1{\kern-0.25pt\vbox{\hrule height 0.2pt\hbox{\vrule
width 0.2pt \kern2pt\vbox{\kern2pt \hbox{#1}\kern2pt}\kern2pt\vrule
width 0.2pt}\hrule height 0.2pt}}

\def\STrow#1{\hbox{#1}\kern-1.35pt}

\font\huge=cmr10 scaled \magstep2
\font\small=cmr8


{\nopagenumbers
\ \ 
\vskip 2cm
\centerline{{\huge\bf Schubert Calculus and Threshold Polynomials of
Affine Fusion}}
\bigskip
\bigskip\bigskip\centerline{S.E. Irvine {\footnote{$^*$}{\small supported
in part by a NSERC Undergraduate Research Award}} \quad\quad M.A. Walton
{\footnote{$^\dagger$}{\small supported in part by NSERC. E-mail:
walton@uleth.ca}}} \bigskip

\centerline{{\it Physics Department,
University of Lethbridge}} \centerline{{\it Lethbridge, Alberta,
Canada\ \ T1K 3M4}}

\vskip 1.5cm \leftskip=1cm \rightskip=1cm
\centerline{{\bf Abstract}}

\noindent We show how the threshold level of affine fusion, the fusion
of Wess-Zumino-Witten (WZW) conformal field theories, fits into the 
Schubert calculus introduced by Gepner. The Pieri rule can be modified
in a simple way to include the threshold level, so that calculations
may be done for all (non-negative integer) levels at once. With the
usual Giambelli formula, the modified Pieri formula deforms the
tensor product coefficients (and the fusion coefficients) into what we
call threshold polynomials. We compare them with the $q$-deformed tensor
product coefficients and fusion coefficients that
are related to $q$-deformed weight multiplicities. We also discuss the
meaning of the threshold level in the context of paths on graphs.  

\leftskip=0cm \rightskip=0cm

\vfill

\eject}

\pageno=1
\newsec{\bf Introduction}

Gepner found geometrical and topological interpretations of the
fusion rings of Wess-Zumino-Witten (WZW) conformal field theories
\ref\gep{D. Gepner,
Commun. Math. Phys. {\bf 141} (1990) 381}. He  described them using a
Schubert calculus, a ``quantum version'' of the
classical Schubert calculus that is fundamental in the geometry
and topology of complex manifolds (see \ref\clSc{H. Hiller, Geometry of
Coxeter groups (Pitman, 1982);\hfb
W. Fulton, Young tableaux (Cambridge University Press, 1997), Part
III;\hfb
P. Griffiths, J. Harris, Principles of algebraic geometry (Wiley,
1978)}, e.g.). 

Gepner also pointed out a correspondence between the WZW fusion rings
and the chiral rings of $N=2$ superconformal theories. These two
observations have been seminal. For example, their relation was
clarified in \ref\witten{E. Witten, in: Geometry,
Topology and Physics; Conf. Proc. and Lecture Notes in
Geom. Topol. {\bf VI} (1995) 357}, where the new Schubert calculus was
shown to describe the quantum cohomology of Grassmannians. Also, the
$N=2$ interpretation led to new realisations of WZW fusion rings in
topological theories 
\ref\intril{K. Intriligator, Mod. Phys. Lett. {\bf A6} (1991)
3543}\ref\vafa{C. Vafa, in: Essays on Mirror Manifolds, ed. S.-T. Yau
(International Press, 1992)}\ref\newwar{D. Nemeschansky, N. Warner,
Nucl. Phys. {\bf B380} (1992) 241}. 

We study the Schubert calculus of WZW fusion rings. Our initial 
motivation was computational. In Gepner's approach, a
fusion potential is introduced whose derivatives give the fusion
constraints to be implemented. The fusion potential, and so the
fusion constraints, are level dependent. Therefore a significant part of any
computation must be re-done whenever the level is changed. By the depth rule
\ref\gepwit{D. Gepner,  E. Witten,
Nucl. Phys. {\bf B278} (1986) 493}, however, the
results are simpler than this procedure indicates. A {\it
threshold level} exists for each coupling \ref\cmw{C.J. Cummins,
P. Mathieu, M.A. Walton, Phys. Lett. {\bf 254B} (1991)
386}\ref\kmsw{A.N. Kirillov, P. Mathieu, D. S\'en\'echal, M.A. Walton,
Nucl. Phys. {\bf B391} (1993) 651}; in any fusion product of
two fixed fields, a third appears in the decomposition for all integer
levels greater than or equal to a characteristic
one\footnote{${}^\dagger$}{To the best of our knowledge, however, a
completely rigorous demonstration of the existence of threshold levels
is still lacking.}. Therefore, finding the threshold levels for a
fixed product amounts to finding the fusion rules for all levels at once. 

We show how to incorporate the notion of threshold level into Gepner's
Schubert calculus for WZW fusion. This is done in section 3, after a
review is given in section 2, where the notation is also established.  

Another motivation for this work emerges in section 2: it is
convenient to encode the threshold levels in generating polynomials,
dubbed {\it threshold polynomials}. These are then polynomial
deformations of tensor product coefficients and fusion
coefficients. Similar objects, the quantum group ($q$-)deformations of
tensor product coefficients \ref\schwar{A. Schilling, S.O. Warnaar,
Commun. Math. Phys. {\bf 202} (1999) 359}\ref\kirshi{A.N. Kirillov,
M. Shimozono, A generalization of the Kostka-Foulkes polynomials,
math.QA/9803062}\ref\shiwey{M. Shimozono, J. Weyman,
Graded characters of modules supported by the closure of a nilpotent
conjugacy class,
math.QA/9804036}\ref\lecthi{B. Leclerc,
J-Y. Thibon, Littlewood-Richardson coefficients and Kazhdan-Lusztig
polynomials, math.QA/9809122} and
fusion coefficients \ref\flot{O. Foda, B. Leclerc, M. Okado,
J-Y. Thibon, Ribbon tableaux
and $q$-analogues of fusion rules in WZW conformal field theories,
math.QA/9810008}\ref\schshi{A. Schilling, M. Shimozono, Bosonic
formula for level-restricted paths, math.QA/9812106}  have been
studied previously. Most importantly to us, the
$q$-deformed coefficients are related to the $q$-deformed weight
multiplicities defined by Lusztig \ref\lusz{G. Lusztig, Ast\'erisque
{\bf 101- 102} (1983) 208-229}. In section 4 we compare
our deformations with the $q$-deformations. We show that the new
deformations are
related in a similar way to deformations of the weight
multiplicities, that are natural from the point of view of a
conjectured refinement \kmsw\ref\kmswii{A.N. Kirillov, P. Mathieu, D.
S\'en\'echal, M.A. Walton, in: Group-Theoretical Methods in Physics,
Proceedings of
the XIXth  International Colloquium, Salamanca, Spain, 1992, Vol. 1
(CIEMAT, Madrid, 1993)}\ref\mwcjp{M.A. Walton,
Can. J. Phys. {\bf 72} (1994) 527} of the Gepner-Witten depth rule
\gepwit.

As we have argued, the threshold level has computational advantages
over the use of
fusion potentials, and the relations derived from them. But again, the 
connection with geometry, topology and $N=2$ superconformal theories
was the point of
\gep, not
computation. There the fusion potential played a central role. But one
can't have it both ways: we indicate at the end of section 3 that a
deformed fusion potential that incorporates the threshold levels
cannot be written. Nevertheless, one might 
hope to give the threshold level a somewhat deeper motivation, perhaps
through  
its meaning in the many  different realisations of WZW fusion
rings.  In section 5 we make a very small start in this direction;
we discuss the meaning of the threshold level in the context of paths
on graphs
(see \ref\zub{J.-B. Zuber, Commun. Math. Phys. {\bf 179} (1996)
265}\ref\petzub{V.B. Petkova, J.-B. Zuber, Nucl. Phys. {\bf B463} (1996) 161},
and references therein).  

Section 6 is a short conclusion.

\newsec{\bf WZW fusion, threshold level, and threshold polynomials}

Let us first establish notation. 
For the most part, we restrict attention to the simple Lie algebras
$A_r$ and the affine
algebras $A^{(1)}_r$ that are the untwisted central extensions of their loop
algebras. When the level $k$ is fixed, we denote the affine algebra by
$A_{r,k}$. However, we use a notation that is easily adapted to any untwisted
affine Kac-Moody algebra $X_r^{(1)}$  (or $X_{r,k}$) based on a simple
Lie algebra $X_r$, and expect that such generalisation
is straightforward.

The set of roots of $X_r$ will be
written as $R$, and the set of positive (negative)  
roots as $R_>$ ($R_<$). If $\alpha\in R$ is a root, then the
corresponding co-root is defined as $\alpha^\vee:=
2\alpha/(\alpha,\alpha)$. 

Let
$F=\{\Lambda_1,\ldots,\Lambda_r\}$ denote the set of fundamental
weights of $X_r$, and 
\eqn\Pwt{P\ :=\ \{\, \lambda = \sum_{i=1}^r\lambda_i\Lambda_i\ |\
\lambda_i\in \Z\, \}  }
the set of integral weights. The set of dominant integral weights, 
\eqn\Pge{P_\ge\ :=\ \{\, \lambda = \sum_{i=1}^r\lambda_i\Lambda_i\ |\
\lambda_i\in \Z_\ge\, \}\ ,  }
is the set of highest weights for irreducible integrable
modules of $X_r$. 

Let $M(\la)$ denote an irreducible module of $X_r$, of highest weight
$\la\in P_\ge$. The set of weights of $M(\lambda)$ will be denoted
$P(\lambda)$. 

The irreducible integrable modules of $X_{r,k}$ have highest
weights that project to the following set of dominant weights of $X_r$:
\eqn\Pk{P^k_\ge\ :=\ \{\, \lambda = \sum_{i=1}^r\lambda_i\Lambda_i\ |\
\lambda_i\in \Z_\ge,\ \sum_{i=1}^r\lambda_ia^\vee_i\le k\, \}\ .  }
The $a^\vee_i$ are the
co-marks, defined by $a_0^\vee=1$, and 
\eqn\thal{\theta^\vee\ =\ \theta\ =\ \sum_{i=1}^r a^\vee_i\alpha_i^\vee\ ,}
where $\theta$ ($\theta^\vee$) denotes the highest (co-)root of $X_r$. We normalise
$(\theta,\theta)=2$.  

The Weyl group of $X_r$ will be denoted by $W$, and the shifted action
of $w\in W$ on a weight $\lambda$ by $w.\lambda = w(\lambda+\rho) - \rho$,
where $\rho = \sum_{i=1}^r\Lambda_i = \sum_{\alpha\in R_>}\alpha/2$ is
the Weyl vector.
$W^k$ will indicate the projection of the affine Weyl group, the
Weyl group of $X_{r,k}$, onto
the horizontal weight space, the weight space of $X_r$. $W$ is
generated by the primitive reflections $r_i$, $i=1,\ldots,r$, with
action
\eqn\riW{r_i\lambda\ =\ \lambda\ -\ (\lambda,\alpha_i^\vee)\alpha_i\ }
on any weight $\lambda$. In order to
enlarge $W$ to $W^k$, we adjoin $r_0$ to the generating set. Its
shifted action is 
\eqn\rzWh{r_0.\lambda\ =\ r_\theta.\lambda + (k+x)\theta\ ,}
where $x$ is the dual
Coxeter number of $X_r$. Notice the $k$-dependence of the action of $W^k$ 
on $P$, coming from that of $r_0$.

We write the decomposition of the tensor product of two irreducible
integrable $X_r$-modules as 
\eqn\tpg{M(\la)\otimes M(\mu)\ =\ \bigoplus_{\nu\in P_\ge}\, T_{\la,\mu}^\nu\,
M(\nu)\ \ .}
We will call the $T_{\lambda,\mu}^\nu\in \Z_\ge$ tensor product
coefficients. 
We indicate the affine fusion of two modules of $X_{r,k}$ by writing the
truncated tensor product of the corresponding modules $M(\la)$ and $M(\mu)$
of $X_r$:  
\eqn\fusk{M(\la)\,\otimes_k\, M(\mu)\ =\ \bigoplus_{\nu\in
P_\ge^k}\, {}^{(k)}T_{\la,\mu}^\nu\, M(\nu)\ \ .}
The fusion coefficients obey
\eqn\kkpi{ {}^{(k)}T_{\la,\mu}^\nu\ \le\ {}^{(k+1)}T_{\la,\mu}^\nu\ ,}
and furthermore
\eqn\limT{\lim_{k\rightarrow\infty}\ {}^{(k)}T_{\la,\mu}^\nu\ =\ 
T_{\la,\mu}^\nu\ . }

We can encode the fusion products for all levels by including the
threshold levels $t$ as subscripts in the tensor product decomposition
\cmw\kmsw. If we denote by ${\cal S}_t$ the operator that includes
these subscripts, we can write 
\eqn\tpgth{{\cal S}_t\left[ M(\la)\otimes M(\mu) \right]\ =\ \bigoplus_{\nu\in P_\ge}\bigoplus_{t\in
\Z_\ge}\, T_{\la,\mu}^{\nu(t)}\, M(\nu)_t\ \ .}
Then  
\eqn\TsumT{{}^{(k)}T_{\la,\mu}^\nu\ =\ \sum_{t=0}^k\,
T_{\la,\mu}^{\nu(t)}\ .}
We call the fixed-threshold-level coefficients
$T_{\lambda,\mu}^{\nu(t)}$, the {\it threshold coefficients}. 
For example, we can modify the $A_2$ tensor product to 
\eqn\tpaii{{\cal S}_t\left[ M(1,1)^{\otimes 2} \right]\ =\ M(2,2)_4 \oplus M(3,0)_3 \oplus M(0,3)_3 \oplus 2M(1,1)_{2,3} \oplus M(0,0)_2\ ,}
encoding the corresponding fusions at all levels. Here
$M(a,b):=M(\lambda)$, with $\lambda=a\Lambda_1 +b\Lambda_2$, and 
$p M(a,b)_{t_1, \ldots,t_p} := 
M(a,b)_{t_1} \oplus\cdots \oplus  M(a,b)_{t_p}$.

From the notational point of view, the threshold levels are
unnecessarily large numbers, since 
\eqn\thwts{T_{\la,\mu}^{\nu(t)}\ \not=\ 0\ \ \Rightarrow\ \ t\,\ge\, (\nu,\theta)\ .}
Consequently, one could also define the {\it threshold delay} $d$ by  
\eqn\tdelay{d\ :=\ t\,-(\nu,\theta)\,\ .}
Writing the delays as superscripts, the right-hand side of \tpaii\ is
replaced by 
\eqn\tpaiid{M(2,2)^2 \oplus M(3,0)^0 \oplus M(0,3)^0 \oplus 2M(1,1)^{0,1} \oplus M(0,0)^2\ .}
This is a minor point, so we'll stick to using the threshold
levels. In section 4, however, \tdelay\ will reappear. 

As \tpaii\ makes clear, we need to consider $N$-tuples of threshold
levels. It is convenient to encode them in {\it threshold polynomials}, defined by
\eqn\tpoly{T_{\lambda,\mu}^\nu[\ell]\ :=\ \sum_{t\in\Z_\ge}\, \ell^{\,t}\, T_{\lambda,\mu}^{\nu(t)}\ \ .}
Then  
\eqn\MMPM{{\cal L}^t\left[ M(\lambda)\otimes M(\mu)\right]\ =\ \bigoplus_{\nu\in
P_\ge}\, T_{\lambda,\mu}^\nu[\ell]\,
M(\nu)\ }
is equivalent to \tpgth. For example, the $A_2$ tensor product \tpaii\
is rewritten as 
\eqn\tppaii{{\cal L}^t\left[ M(1,1)^{\otimes 2} \right]\ =\ \ell^4\, M(2,2) \oplus\ell^3\, M(3,0) \oplus \ell^3\,M(0,3) \oplus (\ell^2+\ell^3)\,M(1,1) \oplus \ell^2\,M(0,0)\ .} 
The threshold polynomials can be regarded as deformations of the tensor product coefficients, since 
\eqn\PiT{T_{\la,\mu}^\nu[1]\ =\ T_{\la,\mu}^\nu\ , }
so that the tensor product \tpg\ is recovered when
$\ell=1$. Furthermore, we define the deformation of the fusion coefficient as 
\eqn\lek{{}^{(k)}T_{\la,\mu}^\nu[\ell]\ :=\ \sum_{t=0}^k\, \ell^t 
T_{\la,\mu}^{\nu(t)} \ .}
So ${}^{(k)}T_{\la,\mu}^\nu[\ell]$ is the degree$\le k$ part of the
polynomial $T_{\la,\mu}^\nu[\ell]$, and 
\eqn\PkT{{}^{(k)}T_{\la,\mu}^\nu[1]\ =\ {}^{(k)}T_{\la,\mu}^\nu\ .}
\limT\ is deformed to  
\eqn\limkT{\lim_{k\rightarrow\infty}\ {}^{(k)}T_{\la,\mu}^\nu[\ell]\ =\ 
T_{\la,\mu}^\nu[\ell]\ ,}
by construction. 
 
From \tpoly, we see that the threshold polynomials are the generating
functions for the threshold coefficients. Consequently,
they are related to the generating functions for fusion rules studied
in \cmw, where the threshold level was first introduced (but named
later in \kmsw). For completeness, we indicate the relation here.

The generating function for fusion rules is defined as 
\eqn\frgen{G(L,M,N;d)\ :=\ \sum_{\lambda,\mu,\nu\in P_\ge}\,
\sum_{k=0}^\infty\, {}^{(k)}T_{\lambda,\mu,\nu}\, d^k\,L^\lambda M^\mu
N^\nu\ \ ,}
where the dummy variables $L,M,N$ satisfy $L^\lambda
L^{\lambda'}=L^{\lambda+\lambda'}$, etc. Here
${}^{(k)}T_{\lambda,\mu,\nu} := {}^{(k)}T_{\lambda,\mu}^{C\nu}$, where
$C\nu$ is the highest weight of the module conjugate to
$M(\nu)$. Using \TsumT\ and switching the order of summations, we
arrive at \eqn\ggen{G(L,M,N;d)\ =\ (1-d)^{-1}\, \sum_{\lambda,\mu,\nu\in
P_\ge}\, L^\lambda M^\mu N^\nu\, T_{\lambda,\mu,\nu}[d]\ .}
Here $T_{\lambda,\mu,\nu}[d]=\sum_{t=0}^\infty\, d^t\,
T_{\lambda,\mu}^{C\nu}[d]$; see \lek\ and \limkT. Hence the only difference
between the generating functions
for deformed tensor product coefficients and fusion rules is
$(1-d)^{-1}$, a factor characteristic of the existence of a threshold
level \cmw.

For successive fusions, we need a memory of the threshold levels. For
example, suppose we need to calculate  ${\cal S}_t\left[
M(\phi)\otimes (M(1,1)^{\otimes 2}) \right]$. Then using \tpaii, we would
encounter products like ${\cal S}_t\left[ M(\phi)\otimes M(1,1)_3
\right]$. So \tpgth\ is only
a special case of what we need. Abusing notation slightly, we attach
threshold levels to the factor modules in the tensor products, and write
\eqn\tpths{M(\la)_r\otimes M(\mu)_s\ =\ \bigoplus_{\nu\in
P_\ge}\bigoplus_{t\in \Z_\ge}\, T_{\la(r),\mu(s)}^{\nu(t)}\, M(\nu)_t\ \ .}
For example, we have
\eqn\tptex{M(1,1)_2\otimes M(1,1)_3\ =\ M(2,2)_4\oplus M(3,0)_3\oplus M(0,3)_3 \oplus 2M(1,1)_{3,3} \oplus M(0,0)_3\ .}
\tpgth\ is recovered from \tpths\ by setting $r=(\lambda,\theta)$ and $s=(\mu,\theta)$. 

Again using polynomials to carry the $N$-tuples of threshold levels, we write
\eqn\tpolyrs{{\cal L}^t\left[ M(\la)_r\,\otimes\, M(\mu)_s \right]\ =\ \bigotimes_{\nu\in
P_\ge}\, T_{\la(r),\mu(s)}^{\nu}[\ell]\,  M(\nu)\ .}
Comparing \tptex\ with \tpaii, for example, shows that there is a
simple relation between the coefficients $T_{\la(r),\mu(s)}^{\nu(t)}$
and 
$T_{\la,\mu}^{\nu(t)}$. To write it in polynomial form, we define  
\eqn\cprod{\ell^a\circ \ell^b\ :=\ \ell^{\,{\rm max}\{a,b\}}\ .}
and extend bilinearly (so that two polynomials can be
multiplied). Then we have 
\eqn\rstpoly{T_{\la(r),\mu(s)}^\nu[\ell]\ =\ \ell^r\circ \ell^s\circ\,
T_{\la,\mu}^\nu[\ell]\  .}
We see that the polynomials $T_{\la,\mu}^\nu[\ell]$ are fundamental,
and so we will  concentrate on them henceforth. 

The definition \cprod, however, is natural from the point of view of
threshold polynomials. With it, we can generalise the crossing
symmetry of the tensor product  coefficients,
\eqn\Pcross{\sum_{\zeta\in P_\ge}\,  T_{\la,\mu}^\zeta\, T_{\zeta,\varphi}^\nu\ {=}\ 
\sum_{\zeta\in P_\ge}\, T_{\la,\zeta}^\nu\,  T_{\mu,\varphi}^\zeta\ ,}
 to 
\eqn\Ptcross{\sum_{\zeta\in P_\ge}\,  T_{\la,\mu}^\zeta[\ell]\,\circ\, T_{\zeta,\varphi}^\nu[\ell]\
{=}\  \sum_{\zeta\in P_\ge}\, T_{\la,\zeta}^\nu[\ell]\,\circ\, T_{\mu,\varphi}^\zeta[\ell]\
.}
Furthermore, the crossing symmetry for the fusion coefficients 
\eqn\Pkcross{\sum_{\zeta\in P_\ge^k}\, {}^{(k)}T_{\la,\mu}^\zeta\, {}^{(k)}T_{\zeta,\varphi}^\nu\ {=}\ 
\sum_{\zeta\in P_\ge^k}\, {}^{(k)}T_{\la,\zeta}^\nu\, {}^{(k)}T_{\mu,\varphi}^\zeta\ }
deforms to
\eqn\Ptkcross{\sum_{\zeta\in P_\ge^k}\, {}^{(k)}T_{\la,\mu}^\zeta[\ell]\,\circ\, {}^{(k)}T_{\zeta,\varphi}^\nu[\ell]\ {=}\ 
\sum_{\zeta\in P_\ge^k}\, {}^{(k)}T_{\la,\zeta}^\nu[\ell]\,\circ\, {}^{(k)}T_{\mu,\varphi}^\zeta[\ell]\ .}

\newsec{\bf Schubert calculus, threshold level, and threshold polynomials}

The Schubert calculus is based on the Pieri and Giambelli
formulas. For discussions of them emphasising the geometric context see
\clSc\gep. More relevant to us is their use in representation theory;
consult \ref\fulhar{W. Fulton, J. Harris, Representation theory: a first
course (Springer-Verlag, 1991)}, e.g. 

The Pieri formula is
\eqn\Pieri{T_{\la,\Lambda}^\nu\ =\ \left\{\matrix{1 \ ,&\ \   {\rm if}\  \nu-\la\in
P(\Lambda)\ ;\cr 0\ ,&\ \ \ {\rm otherwise}\ ,\cr}\right.   }
where $\Lambda$ is a fundamental weight, i.e. $\Lambda\in F$. Here we
are specialising to the algebras $A_r$, although the formulas for
other algebras are only slightly more complicated. 
   
Adapted to include threshold polynomials, the Pieri formula simply becomes
\eqn\tPieri{T_{\la,\Lambda}^\nu[\ell]\ =\ \left\{\matrix{\ell^{\,(\la,\theta)}\circ 
\ell^{\,(\nu,\theta)} \ ,&\ \   {\rm if}\  \nu-\la\in
P(\Lambda)\ ;\cr 0\ ,&\ \ \ {\rm otherwise}\ .\cr}\right.   } 
Fundamental monomials
\eqn\Lmu{M(\Lambda^\mu)\ :=\ M(\Lambda_1)^{\otimes\mu_1}\otimes
M(\Lambda_2)^{\otimes \mu_2}\otimes\cdots \otimes M(\Lambda_r)^{\otimes\mu_r}\ =\ \bigotimes_{i=1}^r\, M(\Lambda_i)^{\otimes\mu_i}\ \ , 
}
with all $\Lambda_i\in F$, are easily decomposed using the Pieri
formula \Pieri. The decompositions are triangular in the irreducible 
highest-weight modules $M(\sigma)$,
$\sigma\in P_\ge$:
\eqn\Llaom{M(\Lambda^\la)\ =\ \bigoplus_{P_\ge\ni\sigma\le\lambda}\,
\Omega_{\la,\sigma}\, M(\sigma)\ \ .} Here $\sigma\le\lambda$ means
that $\lambda-\sigma$ is a non-negative integer linear combination of
positive roots, i.e. 
$\lambda-\sigma\in\Z_\ge R_>$. So $\Omega_{\lambda,\sigma}$ is a
triangular matrix.

The polynomial deformation of this, encoding the threshold levels, is
just \eqn\Llatom{{\cal L}^t\left[ M(\Lambda^\la) \right]\ =\ \bigoplus_{P_\ge\ni\sigma\le\lambda}\,
\Omega_{\la,\sigma}[\ell]\, M(\sigma)\ \ . }
Some $A_2$ examples (to be used shortly) will make this clear. We find:
\eqn\xLlaomi{{\cal S}_t\left[ M(\Lambda^{(2,2)}) \right]\ =\ M(2,2)_4\oplus M(3,0)_3\oplus
4M(1,1)_{2,2,2,3} \oplus M(0,3)_3\oplus 2M(0,0)_{1,2}\ \ ,}
or
\eqn\xLlt{\eqalign{{\cal L}^t\left[ M(\Lambda^{(2,2)}) \right]\ =\ \ell^4\,M(2,2)\oplus& \,\ell^3\,M(3,0)\oplus
(\ell^3+3\ell^2)\,M(1,1) \cr \oplus& \,\ell^3\,M(0,3)\oplus
(\ell+\ell^2)\,M(0,0)\ ;\cr}}
and
\eqn\xLlaomii{{\cal S}_t\left[ M(\Lambda^{(1,1)}) \right]\ =\ M(1,1)_{2}\oplus M(0,0)_{1}\ ,\ \ 
{\cal S}_t\left[ M(\Lambda^{(0,0)}) \right]\ =\ M(0,0)_0\ \ ,}
or
\eqn\xLiit{{\cal L}^t\left[ M(\Lambda^{(1,1)}) \right]\ =\ \ell^2\,M(1,1) \oplus
\ell^1\,M(0,0)\ ,\ \ \ {\cal L}^t\left[ M(\Lambda^{(0,0)}) \right]\ =\ M(0,0)\ \ .}

From \Llaom, the highest-weight modules can be expressed as
polynomials in the fundamental
weights: \eqn\Giam{ M(\sigma)\ =\ \bigoplus_{P_\ge\ni\mu\le\sigma}\,
(\Omega^{-1})_{\sigma,\mu}\, M(\Lambda^\mu)\ \ .}
That is, $M(\sigma)$ can be written as a direct sum of fundamental monomials
$M(\Lambda^\mu)$.  This is the Giambelli formula, in the 
non-determinantal form that can be applied to all simple Lie
algebras, not just $A_r$. Notice that the inversion of
$\Omega$ is greatly simplified by its triangularity, and its inverse
is also triangular. A simple $A_2$ example of \Giam\ is
\eqn\Giami{ M(1,1)\ =\ M(\Lambda^{(1,1)})\ominus M(\Lambda^{(0,0)})\ \ .}

The characters of $X_r$ form an algebra with structure constants equal
to the tensor product coefficients. The Giambelli formula \Giam\ gives
rise to a polynomial realisation of
this character algebra. One simply replaces $M(\Lambda^\mu)$ with
$\prod_{i=1}^rx_i^{\mu_i} =: x^\mu$. The resulting polynomial 
\eqn\schub{S_\sigma(x)\ :=\ \sum_{P_\ge\ni\mu\le\sigma}\,
(\Omega^{-1})_{\sigma,\mu}\, x^\mu\ \ ,} 
is known as a Schur polynomial, a  type of Schubert polynomial
\ref\macd{I. G. Macdonald, Notes on Schubert polynomials (Laboratoire
de combinatoire et d'informatique math\'ematique (LACIM), Universit\'e
du Qu\'ebec \`a Montr\'eal, 1991)}. For example, $x_1x_2-1$ is the
Schur polynomial of the $A_2$ module $M(1,1)$, by \Giami. With the
addition and subtraction of polynomials, the character algebra extends
to a ring.   

Is there a useful threshold-level version of the Giambelli formula?
The inverse $\Omega^{-1}[\ell]$ of the matrix $\Omega[\ell]$ in
\Llatom\ has entries that are negative powers of $\ell$. These are
difficult to interpret in the context of threshold level. We  conclude
that the normal, 
$\ell$-independent matrix $\Omega^{-1}$ should be used. We can write
useful formulas for the threshold polynomials in terms of
$\Omega^{-1}$, and its deformed inverse $\Omega[\ell]$:  
\eqn\Tomomom{ T_{\la,\mu}^\nu[\ell]\ =\
\ell^{(\lambda,\theta)}\circ\ell^{(\mu,\theta)}\circ\, 
\sum_{\al,\beta\in P_\ge}\, 
(\Omega^{-1})_{\la,\al}\, (\Omega^{-1})_{\mu,\be}\,
(\Omega)_{\al+\be,\nu}[\ell]\ \ .}
We'll illustrate this formula on the $A_2$ example with 
$M(\lambda)=M(\mu)=M(1,1)$, using the subscript notation. 
First, the required matrix elements of $\Omega^{-1}$ are provided by 
\eqn\xTomomomi{\eqalign{ M(1,1)^{\otimes 2}\ =&\ 
\left[\, M(\Lambda^{(1,1)})\ominus 
M(\Lambda^{(0,0)})\,\right]^{\otimes 2}\ \cr
=&\ M( \Lambda^{(2,2)})\ominus 2
M(\Lambda^{(1,1)})\oplus M(\Lambda^{(0,0)})\ \ .\cr
}}
Substituting the fundamental monomials \xLlaomi\ and \xLlaomii, described by $\Omega[\ell]$, we get 
\eqn\xii{\eqalign{{\cal L}^t\left[ M(1,1)^{\otimes 2}
\right]\ =&\ \ell^2\circ \bigg\{\ \ell^4  M(2,2)\oplus
\ell^3 M(3,0)\oplus (3\ell^2+\ell^3) M(1,1) \oplus \ell^3
M(0,3) \cr &\ \ \oplus (\ell+\ell^2) M(0,0) \ominus 2\,\big[ \ell^2
M(1,1)\oplus\ell M(0,0)\big]\,
\oplus M(0,0)\ \bigg\}\cr
=&\ \ell^4 M(2,2)\oplus \ell^3 M(3,0)\oplus \ell^3 M(0,3)\oplus
(\ell^2+\ell^3) M(1,1)\oplus \ell^2 M(0,0)\ \ ,\
\cr
 }}
the correct result. 

The deformed Pieri formula \tPieri\ makes straightforward  the
calculation of decompositions involving fundamental monomials, like
$M(\Lambda^\beta)$. We write\eqn\Tfmon{{\cal L}^t\left[
M(\la)\,\otimes\,M(\Lambda^\beta) \right]\
=\ \bigoplus_{\nu\in
P_\ge}\, T^\nu_{\la,\Lambda^\beta}[\ell]\, M(\nu)\ .} 
Then the threshold polynomials may also be calculated from the 
simpler polynomials $T^\nu_{\la,\Lambda^\beta}[\ell]$:   
\eqn\TomT{ T_{\la,\mu}^\nu[\ell]\ =\ \ell^{(\mu,\theta)}\,\circ\,
\sum_{\beta\in P_\ge}\, 
(\Omega^{-1})_{\mu,\be}\, T_{\la,{\Lambda^\beta}}^\nu[\ell]\ \ .}
Using \Giami, an $A_2$ example is
\eqn\xTomT{\eqalign{ {\cal L}^t\left[ M(1,1)^{\otimes 2} \right]\ =&\
\ell^2\circ {\cal L}^t\bigg[\, M(1,1)\otimes \big(
M(\Lambda^{(1,1)})\ominus M(\Lambda^{(0,0)})\big)\,\bigg]\ \cr
=&\ \ \ell^4 M(2,2)\oplus \ell^3 M(3,0)\oplus (2\ell^2+\ell^3)
M(1,1)\oplus \ell^3 M(0,3)\oplus
\ell^2 M(1,1)\cr 
\ &\ominus  \ell^2 M(1,1) \ \ .\cr }}
This is again the correct result (see \xii). 

Finally, we can also multiply two Schur polynomials for $M(\la)$ and
$M(\mu$) together, using the coefficients
$T_{\Lambda^\alpha,\Lambda^\beta}^\nu$, defined in the obvious way: 
\eqn\TomomT{ T_{\la,\mu}^\nu[\ell]\ =\
\ell^{(\la,\theta)}\circ\ell^{(\mu,\theta)} \circ\,\sum_{\al,\beta\in P_\ge}
\, 
(\Omega^{-1})_{\la,\al}\,
\, 
(\Omega^{-1})_{\mu,\be}\, T_{{\Lambda^\al},{\Lambda^\beta}}^\nu[\ell]\
\ .}

To conclude this section, we note that in our deformed Schubert
calculus, there is no analogue of the fusion
potential that was so important in  \gep. We argue that a deformed
potential that incorporates the threshold levels cannot be
written. Gepner could write a fusion potential because at fixed level
$k$, the fusion rules are truncations of the tensor product rules. The
truncated parts can be set to zero by fusion constraints, that can be
derived from the potential. On the other hand, when the threshold
level is incorporated into a tensor product, as in \Llatom\ vs. \Llaom,
there is no truncation. Instead of constraints, one could only hope to
find replacements that would change the right-hand side of \Llaom\
into that of \Llatom, for example. But that is exactly what we do: \Llatom\
is obtained from \Llaom\ by replacing $\Omega$ with $\Omega[\ell]$. A
minimal set of such replacements would be those obtained by replacing
the right-hand side of the undeformed Pieri rule \Pieri\ with that of
the deformed one \tPieri.  

Incidentally, we have seen that the Pieri rule with
threshold level \tPieri\ contains the same information as the fusion
potentials of Gepner, for all (non-negative integer) levels. So does 
the generating function for the fusion potentials \gep. It might
be interesting to make this more precise.

\newsec{\bf Deformed tensor product coefficients and weight multiplicities}

The threshold polynomials \tpoly\ and \lek\ are deformations of the
tensor product coefficients and affine (WZW) fusion coefficients,
respectively. It is interesting to compare them with the
$q$-deformations of these objects studied previously.

WZW fusion coefficients are alternating
affine-Weyl ($W^k$) 
sums of tensor product coefficients \ref\kac{V.G. Kac, Infinite 
dimensional Lie algebras, 3rd ed.
(Cambridge U. Press, 1990)}\ref\mwrs{M.A. Walton, Phys. Lett. {\bf
241B} (1990) 365; Nucl. Phys. {\bf B340} (1990)
777}\ref\fgp{P. Furlan, A. Ganchev, 
V. Petkova,
Nucl. Phys. {\bf B343} (1990) 205}\ref\goowen{F. Goodman, H. Wenzl,
Adv. Math. {\bf 82} (1990) 244}. In \flot\ (see also \schshi), the
corresponding $q$-fusion coefficients (for affine $A_r$) are defined
in similar fashion in terms of the
$q$-tensor product coefficients \schwar\kirshi\shiwey\lecthi. 
Since the ordinary (undeformed) tensor product coefficients are also
alternating Weyl sums of the weight multiplicities of Lie algebras,
the fusion coefficients can also be expressed in that way. In the
$q$-deformed case, the tensor product and fusion coefficients are
related to Lusztig's $q$-deformed weight multiplicities
\lusz, in turn related to the famous Kazhdan-Lusztig polynomials
\ref\klpoly{D. Kazhdan, G. Lusztig, Invent. Math. {\bf 53} (1979)
165}.

Let us start with an example, taken from \flot. They find, for the
$q$-deformation of the $A_3$ tensor product $M(1,1,0)^{\otimes 3}$, the
following decomposition:     
\eqn\flotex{\eqalign{(q^3+&q^6)\,M(0,0,3) \oplus
(2q^4+3q^5+2q^6+q^7)\,M(0,1,1) \oplus (q^2+2q^3+q^4)\, M(0,3,1)\cr
    &\oplus (q^5+2q^6+q^7)\, M(1,0,0) \oplus (q^2+2q^3+3q^4+2q^5)\,
    M(1,1,2)\ \cr &\oplus (2q^3+3q^4+3q^5+q^6)\, M(1,2,0) \oplus
    (q+q^2)\, M(1,4,0) \cr &\oplus (q^3+3q^4+3q^5+2q^6)\, M(2,0,1)
    \oplus (q+2q^2+2q^3+q^4)\, M(2,2,1)\cr &\oplus (q^2+2q^3+q^4)\,
    M(3,0,2) \oplus (q^2+2q^3+2q^4+q^5)\, M(3,1,0)\cr &\oplus
    \,M(3,3,0) \oplus (q+q^2)\, M(4,1,1) \oplus (q^3)\, M(5,0,0)\ .\cr}} 
This should be compared with the threshold-level version of the same 
tensor product:
\eqn\flottl{\eqalign{{\cal L}^t\left[ M(1,1,0)^{\otimes 3} \right]\ =\
    &(2\ell^3)\,M(0,0,3) \oplus
(\ell^2+7\ell^3)\,M(0,1,1) \oplus (4\ell^4)\, M(0,3,1)\cr 
    &\oplus (2\ell^2+2\ell^3)\, M(1,0,0) \oplus (8\ell^4)\,
    M(1,1,2)\ \cr &\oplus (4\ell^3+5\ell^5)\, M(1,2,0) \oplus
    (2\ell^5)\, M(1,4,0) \cr &\oplus (5\ell^3+4\ell^4)\, M(2,0,1)
    \oplus (6\ell^5)\, M(2,2,1)\cr &\oplus (4\ell^5)\,
    M(3,0,2) \oplus (4\ell^4+2\ell^5)\, M(3,1,0)\cr &\oplus (\ell^6)
    \,M(3,3,0) \oplus (2\ell^6)\, M(4,1,1) \oplus (\ell^5)\, M(5,0,0)\
    .\cr}}
From this example, we see no clear relation between the
    $q$-deformations and the $\ell$-deformations, except that they
    coincide at $q=\ell=1$. 

In order to define the $q$-tensor product coefficients, one introduces
the $q$-deformed Kostant partition function $K(\beta;q)$:
\eqn\qkos{\prod_{\alpha\in R_>}\, (1-qe^\alpha)^{-1}\ =:\
\sum_{\beta\in\Z_\ge R_>}K(\beta;q)e^\beta\ \ .} 
From this we see that the powers of $q$ count the number of positive
roots in a decomposition of an element of $\Z_\ge R_>$. 
The $q$-deformed weight multiplicities are 
\eqn\qmult{{\rm mult}_\lambda(\mu;q)\ :=\ \sum_{w\in W}\, (\det w)\,
K(w.\lambda-\mu;q)\ ,}
and we get the 
$q$-deformed tensor product coefficients as 
\eqn\qtp{T_{\lambda,\mu}^\nu(q)\ :=\ \sum_{w\in W}\, (\det w)\, {\rm
mult}_\mu(w.\nu-\lambda;q)\ .}
Notice we use different brackets to distinguish the different
deformations: $T_{\lambda,\mu}^\nu(q)$
vs. $T_{\lambda,\mu}^\nu[\ell]$. Finally, the $q$-fusion coefficients
\flot\ can be found from 
\eqn\qkTT{{}^{(k)}T_{\lambda,\mu}^\nu(q)\ :=\ \sum_{w\in W^k}\,
(\det w)\, T_{\lambda,\mu}^{w.\nu}(q)\ .}

It would be interesting to define the threshold polynomial versions of
the $q$-Kostant partition function, and  the $q$-multiplicities. The
relation between the $q$-tensor product coefficients and the threshold
polynomials might then be extracted. We have not succeeded in finding
the ``$\ell$-Kostant partition function''. But the
$\ell$-multiplicities may be defined using a conjectural refinement
\kmsw\kmswii\ of the Gepner-Witten depth rule \gepwit:
\eqn\rdr{{}^{(k)}T_{\lambda,\mu}^\nu\ =\ \dim\left\{\, v\in
M(\mu;\nu-\lambda)\,|\, (f_i)^{\nu_i+1} v =0,\, \forall
i\in\{0,1,\ldots,r\}\, \right\} \ \ .}
Here $M(\mu;\nu-\lambda)$ is the subspace  of weight $\nu-\lambda$ of
the module $M(\mu)$, so that $\dim M(\mu;\sigma) = {\rm
mult}_\mu(\sigma)$. The $f_i$ are the lowering operators corresponding
to the simple roots of 
the simple Lie algebra $X_r\subset X_{r,k}$, for $i=1,\ldots,r$. $f_0$
is identified with $e_\theta$, the raising operator corresponding to
the highest root $\theta$ of $X_r$. Recall that the {\it depth} of a
vector $v$ (in a module $M(\mu)$, say) is defined as the non-negative
integer $d$ such that $(e_\theta)^{\,d}\, v \not=0$, while
$(e_\theta)^{\,d+1}\, v =0$. The relation between \rdr\ and the
Gepner-Witten depth rule is then clear. 

By \rdr, we see that the subspaces 
\eqn\Mmusid{M(\mu;\sigma | d)\ :=\ \{\, v\in M(\mu;\sigma)\ |\ 
(e_\theta)^{\,1+d}\, v =0, \ (e_\theta)^{d}\, v\not =0\,\}\  }
of the spaces
$M(\mu;\sigma)$ are relevant.
Because of their relation to the depth, the multiplicities that are
the dimensions of these 
spaces,
\eqn\prof{{\rm mult}_\mu(\sigma | d)\ :=\ \dim
  M(\mu; \sigma | d)\ ,}   
were dubbed ``profundities'' in \mwcjp. They have the same relation to the
threshold coefficients (see \TsumT) that the the usual multiplicities
have to the tensor product coefficients:
\eqn\tcprof{T_{\lambda,\mu}^{\nu(t)}\ =\ \sum_{w\in W}\, (\det w)\,
{\rm mult}_\mu\left( w.\nu-\lambda\, |\, t-(\nu,\theta)\right)\ ,}
where $t\ge(\nu, \theta)$. Substituting this into \tpoly\ gives 
\eqn\ltp{T_{\lambda,\mu}^\nu[\ell]\ :=\
\ell^{\,(\nu,\theta)}\,\sum_{w\in W}\, (\det w)\, {\rm
mult}_\mu(w.\nu-\lambda;\ell)\ ,}
with $\ell$-deformed multiplicities
\eqn\lmult{{\rm
mult}_\mu(\sigma;\ell)\ :=\ \sum_{d\in \Z_\ge}\,\ell^{\,d}\, {\rm
mult}_\mu(\sigma | d)\ .}
So the only complication is the overall factor
$\ell^{\,(\nu,\theta)}$, and the $\ell$-deformed
multiplicities are generating functions for the profundities. 

Incidentally, in deriving this last result, we used the relation 
\eqn\dtnuth{d\ =\ t\ -\ (\nu,\theta) }
between the depth $d$ and the threshold level $t$ of a fixed
``coupling''. Notice that this is identical to \tdelay. Hence the
threshold delay of a coupling equals the depth of the corresponding
vector in \rdr.

\newsec{\bf Fusion paths and the threshold level}

Paths on fusion graphs are important in certain integrable lattice
models that are related to conformal field theory (see \zub\petzub, and
references therein). These graphs may also have a more fundamental
significance, indicated by the correspondence between $A_1$
modular invariants and the $A-D-E$ graphs \ref\ciz{A. Cappelli, C. Itzykson,
J.-B. Zuber, Commun. Math. Phys. {\bf 113} (1987) 1}. 

If we restrict to the case $X_{r,k}=A_{r,k}$, then the points of the relevant
graphs correspond to the weights of $P_\ge^k$. There is a distinct
directed 
graph ${}^{(k)}{\cal G}_i$ for each fundamental weight
$\Lambda_i\in F$. The edges of the graph ${}^{(k)}{\cal G}_i$ are
encoded in its incidence
matrix ${}^{(k)}G_i$, which is not necessarily
symmetric. $({}^{(k)}G_i)_{\lambda,\mu}$ is the
number of edges joining node $\lambda$ with node $\mu$. The fusion graph is
defined by $({}^{(k)}G_i)_{\lambda,\mu}=
{}^{(k)}T_{\Lambda_i,\lambda}^\nu$, or ${}^{(k)}G_i =
{}^{(k)}T_{\Lambda_i}$, hence the name. One can also define a graph
${}^{(k)}{\cal G}$ with incidence matrix ${}^{(k)}G:= \sum_{i=1}^r
{}^{(k)}G_i$.  

A fusion path is a path on a fusion graph. Such paths parametrise the
Hilbert space of certain integrable two-dimensional lattice
models. The basic construction is a representation of a
(quotient of a) Hecke algebra on this space. It guarantees
that the models' Boltzmann weights satisfy the Yang-Baxter
equation, ensuring integrability. 

Due to \kkpi,\limT, and since $P^k_\ge
\subset P^{k+1}_\ge \subset \cdots\subset 
P_\ge$, we can think of the graphs ${}^{(k)}{\cal G}_i$ and
${}^{(k)}{\cal G}$ in the
infinite-level limit as tensor product graphs. (Here we restrict
consideration to paths involving weights that do not increase with the
level.) Such paths on $P_\ge$ will also be paths on all ${}^{(k)}{\cal
G}$, for all levels $k$ greater than a certain threshold level
$t$. This threshold level, is just the maximum height
$ht(\lambda):=(\lambda,\theta)$ of the weights $\lambda\in P_\ge$ on
the path.  

Key to the modified Schubert calculus described above were the
fundamental monomials, and their decompositions \Llaom. But the fundamental
monomials $M(\Lambda^\mu)$ 
generate paths in $P_\ge$: to every module $M(\sigma)$ in the
decomposition \Llaom\ there corresponds a path on $P_\ge$ that begins
at the weight $0$ and ends at $\lambda$. To each factor
$M(\Lambda)$, $\Lambda\in F$,  in the monomial corresponds a segment
of the path that
connects nodes of the graph that differ by some $\varphi\in
P(\Lambda)$. 

The threshold level is included by modifying \Llaom\ to \Llatom, using
the $\ell$-Pieri rule. From \tPieri, we see that
the threshold level is the
maximum height of a path on the
(infinite)  tensor product
graph of $A_r$. This is the main point of this section. 

We should emphasise, however, that the correspondence between the
fundamental monomials and
tensor product paths is not one-to-one. The  
polynomial realisation \schub\ of the fusion ring is possible because
$M(\Lambda)\otimes M(\Lambda') = M(\Lambda')\otimes M(\Lambda)$, for
all $\Lambda,\Lambda'\in F$. But the order of tensor product factors
$M(\Lambda)$ 
changes the path. This can be made clear by writing a 
generating matrix for fundamental monomials: 
\eqn\gfmon{\Phi\ :=\ \sum_{\mu\in P_\ge}\, e^\mu\,
T_{\Lambda_1}^{\mu_1} \cdots T_{\Lambda_r}^{\mu_r}\ =\ \prod_{i=1}^r\,
\left[ 1- e^{\Lambda_i}\,T_{\Lambda_i} \right]^{-1}\ .}
Here $e^\mu$ denotes a formal exponential, satisfying $e^\mu e^\nu =
e^{\mu+\nu}$. $\Phi_{\lambda,\mu}$ will equal the sum over all
fundamental monomials that when tensored with $M(\lambda)$, include
$M(\mu)$ in the decomposition, multiplied by the formal exponential of
the  monomial weight of each. Putting $e^\mu \rightarrow 1$ then gives
the number of such monomials. On the other
hand, to generate all paths connecting nodes $\lambda$ and  $\mu$, we
need $\Theta_{\lambda,\mu}$ instead, where
\eqn\gfpath{\Theta\ :=\ \left[\, 1- \sum_{i=1}^r
e^{\Lambda_i}T_{\Lambda_i} \,\right]^{-1}\ .}

The deformations of these two generating matrices are simple to
write. One only needs to replace the tensor product matrices
$T_{\Lambda_i}$ with their $\ell$-deformations, and insist that they
are multiplied in the manner of \Ptcross. So we get
\eqn\gfmonl{\Phi[\ell]\ =\ \prod_{i=1}^r\,
\left\{ 1- e^{\Lambda_i}\,T_{\Lambda_i}[\ell] \right\}^{(\circ\,-1)}\ ,}
where the notation (we hope) is clear, and a similar formula analogous
to \gfpath.

\newsec{\bf Conclusion}

Our main result is a Schubert-type calculus for affine fusion that
incorporates the threshold level. At fixed level, fusion constraints are
natural because a  fusion rule is
a truncation of a tensor product decomposition. Thus fusion potentials
that generate constraints are possible, if not necessary. On the other
hand, in the
threshold level formalism one doesn't truncate a tensor product, but
rather replaces it with a deformed version. So, instead
of using a fusion potential to generate constraints, one just deforms
the tensor  products and then all (non-negative integer) levels are
treated on equal footing. The
deformations are generated by the deformed version of the Pieri rule,
\tPieri.  

In summary then, to include  the threshold levels in a calculus of Schubert
type, use the  undeformed
Giambelli formula \Giam,  and the deformed Pieri formula \tPieri. Then
the threshold polynomials can be calculated by \Tomomom,\TomT, or
\TomomT. 

Another result is the comparison in section 4 of the threshold
polynomials with the $q$-deformed tensor product and fusion
coefficients. In particular, we found the analogue of the $q$-deformed
weight multiplicities in our $\ell$-deformation.  

We also discussed the interpretation of the threshold level for the
decomposition of fundamental monomials, as in \Llatom. In the
corresponding path on a tensor product graph, the threshold level is
just the maximum height of weights on that path. 

To close, let us mention a few possible directions from this work. 

One could hope to make the connection between the $q$-deformations and
$\ell$-deformations more precise, extending our section 4. We also
expect that one could define a $q$-Schubert calculus for the
$q$-tensor product coefficients \qtp, in a straightforward way. In
contrast with the $\ell$-deformed case, both $\Omega(q)$ and
$\Omega^{-1}(q)$ should be important. It might be of interest to
introduce $q$-analogues of the fusion
constraints and potentials of Gepner's calculus, for the
$q$-fusion-coefficients.  

The $\ell$-deformed Schubert calculus is relevant to the search
for a Littlewood-Richardson rule for affine fusion
\ref\mwdem{M.A. Walton, J. Math. Phys. {\bf 39} (1998) 665}. In the
present context, the usual Littlewood-Richardson rule for tensor
products is related to \TomT, at $\ell=1$. This formula involves the
tensor product of two modules $M(\lambda),\, M(\mu)$, where one is
expressed in terms of fundamental monomials by \Giam: $M(\mu)=\sum_\beta 
(\Omega)^{-1}_{\mu,\beta} M(\Lambda^\beta)$. The rule gives a way of
avoiding the cancellations inherent in \TomT\ (see \xii, e.g.). It
identifies a choice of a part of the decompositions of the
$M(\Lambda^\beta)$ in \TomT\ that leads directly to the result. Unfortunately,
the deformed Pieri rule applied to that choice 
gives incorrect threshold levels (one finds $2M(1,1)_{2,2}$ instead of
the $2M(1,1)_{2,3}$ of \xii, e.g.). Calculations of the type \TomT,
however, show us all the parts. And so we can hope that more in-depth analysis
will reveal the appropriate modification.

Finally, if it exists, a motivation other than computational for the
threshold level should be found. It might be revealed by finding
the meaning of the threshold level in the many different physical and
mathematical realisations of affine fusion.

\vskip 2truecm
\noindent{\it Acknowledgements}

We thank L. B\'egin, T. Gannon, P. Mathieu and J. Rasmussen for comments, and
D. S\'en\'echal for the use of some computer programs.

\listrefs
\bye